\def\addr#1{{\small\it #1}}

\def\etal{{\frenchspacing\it et al.}}
\def\ie{{\frenchspacing\it i.e.}}
\def\eg{{\frenchspacing\it e.g.}}

\def\beq#1{\begin{equation}\label{#1}}
\def\eeq{\end{equation}}
\def\beqa#1{\begin{eqnarray}\label{#1}}
\def\eeqa{\end{eqnarray}}
\def\eq#1{equation~(\ref{#1})}

\def\eqnum#1{~(\ref{#1})}

\def\bfig{\begin{figure}[h] \centerline{\hbox{}}\vfill}
\def\efig{\end{figure}\vfill\newpage}

\def\pp{\noindent\parshape 2 0truecm 13.6truecm 1truecm 12.6truecm}
\def\rf#1;#2;#3;#4 {\par\pp#1, {\it #2}, {\bf #3}, #4. \par}
\def\rg#1;#2;#3;#4;#5 {\par\pp#1, {\it #2}, {\bf #3}, #4 (``#5"). 
\par}
\def\rn{\pp}

\def\tento#1{\times10^{#1}}
\def\aet#1#2{\approx #1 \tento{#2}}

\def\Ms{M_{\odot}}     

\def\K{{\rm K}}
\def\s{{\rm s}}

\def\cm{{\rm cm}}
\def\eV{{\rm eV}}
\def\Mpc{{\rm Mpc}}
\def\kpc{{\rm kpc}}

\def\erfc{{\rm erfc}}
\def\st{\sigma_t} 
\def\Ob{\Omega_b}  
\def\Oz{\Omega_0}
\def\soz{\sqrt{1+\Oz z}}

\def\etal{{\frenchspacing\it et al.}}
\def\ie{{\frenchspacing\it i.e.}}
\def\eg{{\frenchspacing\it e.g.}}
\def\izi{\int_0^{\infty}}

\def\izz{\int_0^z}

\def\expec#1{\langle#1\rangle}

\def\ps{P_s}
\def\i{x}        % Ionization
\def\v{\chi}     % Volume fraction ionized

\def\crr{\cr\noalign{\vskip 4pt}}

\def\fuv{f_{uv}}
\def\fupp{f_{uvpp}}
\def\euv{\expec{E_{uv}}}

\def\uvsigfid{\sigma_{18}}

\def\lrec{\lambda_{rec}}  
\def\lci{\lambda_{ci}}
\def\lpi{\lambda_{pi}}
\def\lcomp{\lambda_{comp}}

\def\fnet{f_{net}}
\def\fH{f_H}
\def\fs{f_s}

\def\fmet{f_{burn}}
\def\fbh{f_{bh}}
\def\facc{f_{acc}}
\def\fuv{f_{uv}}
\def\fion{f_{ion}}
\def\fesc{f_{esc}}
\def\szmc{\sigma(M_c,0)}
\def\zvir{z_{vir}}
\def\zion{z_{ion}}
\def\opzopzv{\left({1+z\over 1+\zvir}\right)}

\def\Tpi{T^*}
\def\ktmc{\left({kT\over m_e c^2}\right)}
\def\recfac{\eta_{rec}}

\documentstyle[11pt,epsf,rotate]{article}

\begin{document}

%%%%%%%%%%%%%%%%%%%%%%%%%%%%%

\begin{titlepage}   % Not numbered.

\noindent
%\today
%August 27, 1996
%\hfill MPI-PhT/96-81
\begin{center}

\vskip0.9truecm
{\bf

ON THE INEVITABILITY OF REIONIZATION:\\
IMPLICATIONS FOR COSMIC MICROWAVE BACKGROUND FLUCTUATIONS\footnote{
Published in {\it ApJ}, {\bf 420}, 486, January 10, 1994.\\
Submitted March 18 1993, accepted July 2.
Available from\\
{\it h t t p://www.sns.ias.edu/$\tilde{~}$max/reion.html} 
(faster from the US) and from\\
{\it h t t p://www.mpa-garching.mpg.de/$\tilde{~}$max/reion.html} 
(faster from Europe).\\
}
}

\vskip 0.5truecm

Max Tegmark$^1$, 
Joseph Silk$^2$ 
\&
Alain Blanchard$^3$

\smallskip
\addr{$^1$Department of Physics, University of California, 
Berkeley, California  94720}\\
\addr{$^2$Departments of Astronomy and Physics, and
Center for Particle Astrophysics, University of California, 
Berkeley, California 94720}\\
\addr{$^3$DAEC, Observatoire de Paris-Meudon, 92190 Meudon, France}\\

\smallskip
\vskip 0.2truecm

\end{center}

\begin{abstract}
Early photoionization of the intergalactic medium is discussed in a
nearly model-independent way, in order to 
investigate whether early structures
corresponding to rare Gaussian peaks in a CDM model can
photoionize the intergalactic medium sufficiently 
early to appreciably smooth out the microwave background fluctuations.
We conclude that this is indeed possible for a broad range
of CDM normalizations and is almost inevitable for unbiased CDM, 
provided that the
bulk of these early structures are quite small, no more massive than about
$10^8 M_{\odot}$. Typical parameter values predict that reionization
occurs around $z=50$, thereby suppressing fluctuations on degree scales
while leaving the larger angular scales probed by COBE reasonably unaffected.
However, for non-standard CDM, incorporating mixed dark matter,
vacuum density or a tilted primordial power spectrum, early reionization
plays no significant role.
\end{abstract}
\end{titlepage}
%%%%%%%%%%%%%%%%%%%%%%%%%%%%%%%%%%%%%%%%%%%%%%

\section{Introduction}

The first quantitative predictions of cosmic microwave background
anisotropies in cold dark matter (CDM)-dominated cosmological models
recognized that reionization by rare, early-forming objects could play a
role in suppressing temperature fluctuations on small angular scales (Bond
\& Efstathiou 1984; Vittorio \& Silk 1984). Now that the COBE DMR
experiment has detected fluctuations on large angular scales
(Smoot {\etal} 1992) at a level
(within a factor of two) comparable to that predicted  by CDM models,
it is especially relevant to examine whether reionization can affect the
degree scale searches that are currently underway.

Cold dark matter models are generally characterized by a late epoch of
galaxy formation. However, the smallest and oldest objects first go
nonlinear at relatively large redshift. In this paper we investigate,
for a wide range of CDM normalizations, power spectra and
efficiency parameters, whether reionization associated with energy
injection by early forming dwarf galaxies can reionize the universe
sufficiently early to smooth out primordial CBR temperature
fluctuations.  

Although we go into some detail in the appendix
to make estimates of a certain efficiency parameter, our overall
treatment is fairly model-independent, and can be used as a
framework within which to compare various photoionization scenarios.
Our basic picture is roughly the following: 
An ever larger
fraction of the baryons in the universe falls into nonlinear structures
and forms galaxies. A certain fraction of these baryons form
stars or quasars which emit ultraviolet radiation, 
and some of this radiation escapes into the ambient 
intergalactic medium (IGM) and ends up
photoionizing and heating it. Due to cooling losses and
recombinations, the net number of ionizations per UV
photon is generally less than unity. 

Apart from photoionization, early galaxies can also ionize 
the IGM through
supernova driven winds, an ionization mechanism that will not be treated
in this paper. Although such winds can ionize the IGM by $z=5$, early
enough to explain the absence of a Gunn-Peterson effect (Tegmark {\etal}
1993), the relatively low velocities of such winds makes them unable to
distribute the released energy throughout space at redshifts early enough
(by $z\approx 50$) to measurably affect the CBR.

Our approach will be to first write the ionization fraction of the
IGM as a product of a number of factors, and then discuss
the value of each of these factors in more detail. Let us
write
\beq{FirstFactorEq}
\v = \fs\fupp\fion,
\eeq
where
$$\cases{
\v&= fraction of IGM that is ionized,\cr
\fs&= fraction of baryons in nonlinear structures,\cr
\fupp&= UV photons emitted into IGM per proton in nonlinear structures,\cr
\fion&= net ionizations per emitted UV photon.
}$$
Let us first consider the case where the UV photons are produced by
stars,
and return to the quasar case later.
Using the fact that a
fraction $0.0073$ of the rest mass is released in stellar burning of
hydrogen to helium, we obtain
\beq{fuppEq}
\fupp \approx 0.0073\left({m_pc^2\over 13.6\eV}\right)
\fH\fmet\fuv\fesc,
\eeq
where
$$\cases{
\fH&= mass fraction hydrogen in IGM,\cr
\fmet&= mass fraction of hydrogen burnt,\cr
\fuv&= fraction of energy released as UV photons,\cr
\fesc&= fraction of UV photons that escape from galaxy.
}$$
We will take the primordial mass fraction of helium to be $24\%$, 
{\ie} $\fH=76\%$.
Now define the {\it net efficiency} 
$$\fnet = \fmet\>\fuv\>\fesc\>\fion,$$
and \eq{FirstFactorEq} 
becomes
\beq{SecondFactorEq}
\v\aet{3.8}{5}\>\fnet\>\fs.
\eeq
The key feature to note about this expression is that
since $3.8\times 10^5$ is such a large number, quite modest 
efficiencies $\fnet$ still allow $\v$ to become of order
unity as soon as a very small fraction of the
baryons are in galaxies. 
As will be seen in the next section, this means
that reionization is possible even at 
redshifts far out in the Gaussian tail of the
distribution of formation redshifts, at epochs long before
those when the bulk of the baryons go nonlinear. This appears to have
been first pointed out by Couchman and Rees (1986).

\section{The Mass Fraction in Galaxies}

In this section, we will discuss the parameter $\fs$.
Assuming the standard PS theory of structure
formation (Press \& Schechter 1974), the fraction of all mass that has
formed gravitationally bound objects of total (baryonic
and non-baryonic) mass greater than $M$ at redshift $z$
is the integral of the Gaussian tail, 
\beq{3fgEq}
\fs = \erfc\left[{\delta_c\over\sqrt{2}\sigma(M,z)}\right],
\eeq
where the complementary error function  $\erfc(x)\equiv
2\pi^{-1/2}\int_x^{\infty} e^{-u^2}du$ and $\sigma(M,z)$ is
the r.m.s. mass fluctuation in a sphere containing an
expected mass $M$ at redshift $z$.
$\sigma^2$ is given by top-hat filtering of the power
spectrum as  
\beq{3FilterEq}
\sigma(M,z)^2 \propto
\izi P(k) 
\left[{\sin kr_0\over(kr_0)^3} - {\cos
kr_0\over(kr_0)^2}\right]^2 dk,
\eeq
where $P(k)$ is the power spectrum at redshift $z$ and 
$r_0$ is
given by ${4\over 3}\pi r_0^3\rho = M$, 
$\rho = {3H^2\Omega\over 8\pi G}$ being
the density of the universe at redshift $z$. Although this approach
has been criticized as too simplistic, numerical
simulations (Efstathiou {\etal} 1988; Efstathiou \&
Rees 1988; Carlberg \& Couchman 1989) have shown that it
describes the mass distribution of newly formed structures
remarkably well. Making the standard assumption of a
Gaussian density field, Blanchard {\it et al.} (1992) have
argued that it is an accurate description at least in the
low mass limit. Since we are mainly interested in
extremely low masses such as $10^6\Ms$, it appears to
suffice for our purposes.

For our middle-of-the road estimate, we choose $\delta_c = 1.69$,
which is the linearly extrapolated overdensity at which a spherically
symmetric perturbation has collapsed into a virialized
object (Gott \& Rees 1975; Efstathiou {\etal} 1988; 
Brainerd \& Villumsen 1992).
We take
$\delta_c = 1.44$ (Carlberg \& Couchman 1989) for the optimistic
estimate, although the even lower value
$\delta_c = 1.33$ has been discussed (Efstathiou \& Rees 1988),
and $\delta_c = 2.00$ (Gelb \&
Bertschinger 1992) for the pessimistic estimate.
(Here and
throughout this paper, parameter choices are referred to as optimistic
if they permit earlier reionization.)

The fact that $\sigma(M,z)\to\infty$ as $M\to 0$ implies
that if we consider arbitrarily small scales, then all
dark matter is in non-linear structures. Thus if
no forces other than gravity were at work, so that
the baryons always followed the dark matter, we would
simply have $\fs=1$ at all $z$. 
However, it is
commonly believed that galaxies correspond only to
objects that are able to cool (and fragment into stars) in
a dynamical time or a Hubble time (Binney 1977; Rees \&
Ostriker 1977; Silk 1977; White \& Rees 1978).
The former applies to ellipticals and bulges, the latter to disks. 
Let us define the {\it virialization redshift} $(1+\zvir)\equiv
(\sqrt{2}/\delta_c)\szmc$,
where $M_c$ is some characteristic cutoff mass which is
the total mass (baryonic and dark) of the first galaxies
to form. $\zvir$ is roughly the
redshift at which the bulk of all baryons 
goes non-linear. 
Using \eq{3fgEq} and the fact that 
$\sigma(M,z) = \sigma(M,0)/(1+z)$ in the linear regime of CDM, 
we thus have 
\beq{fsEq}
\fs = \erfc\left[{1+z\over 1+\zvir}\right].
\eeq
A common assumption is that $M_c\approx
10^6\Ms$, roughly the Jeans mass at recombination. 
Blanchard {\etal} (1992) examine the interplay
between cooling and gravitational collapse in considerable
detail, and conclude that the first galaxies to form have
masses in the range $10^7\Ms$ to  $10^8\Ms$, their redshift
distribution still being given by \eq{fsEq},
whereas Couchman \& Rees (1986) argue that the first
galaxies to form may have had masses as low as $10^5\Ms$. 

As our CDM power spectrum today, we will use that given by BBKS
(Bardeen {\etal} 1986) 
and an $n=1$ Harrison-Zel'dovich primordial spectrum:
$$P(k) \propto 
\left({q^{-1} \ln(1+2.34q)\over
\left[1+3.89q + (16.1q)^2 + (5.46q)^3
+ (6.71q)^4\right]^{1/4}}\right)^2 q,$$
where $q\equiv k/[h^2\Omega_0\Mpc^{-1}]$. Throughout this paper, we
will take $\Omega_0 = 1$. 

\begin{figure}[phbt]
\centerline{\epsfxsize=17cm\epsfbox{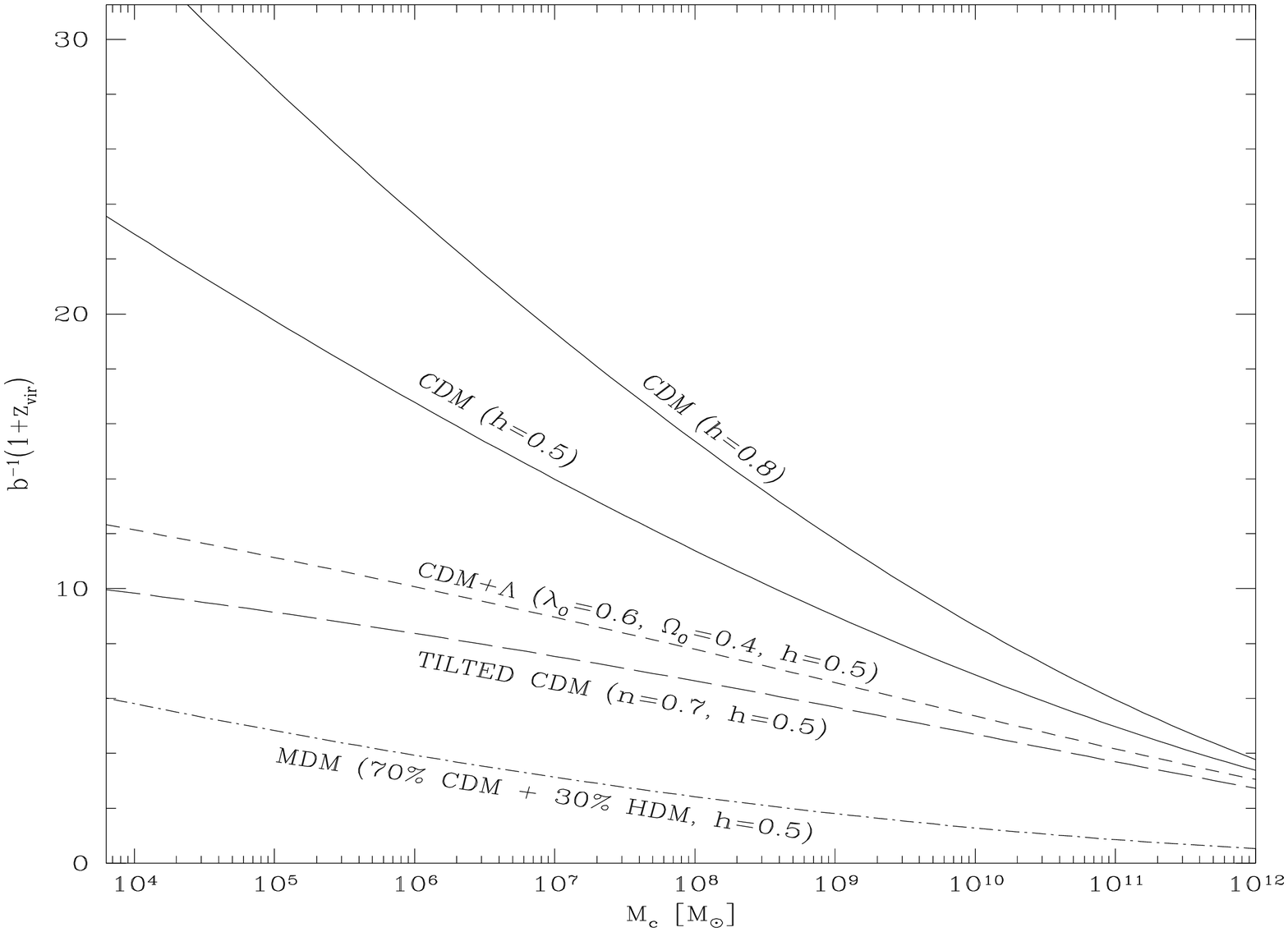}}
\caption{Virialization redshifts for objects of various masses.}
The virialization redshift, the redshift at which the bulk of the 
objects of mass $M_c$ form, is plotted for a number of cosmological 
models. In all cases shown, $\Omega_0+\lambda_0=1$ and $\delta_c = 1.69$.
\label{reionfig1}
\end{figure}

Evaluating the $\sigma^2$-integral in
\eq{3FilterEq} numerically yields 
$$\sigma(10^5\Ms,0)\approx 33.7b^{-1}$$
for $h=0.8$ and 
$$\sigma(10^8\Ms,0)\approx 13.6b^{-1}$$
 for $h=0.5$, where the so
called bias factor $b\equiv\sigma(8h^{-1}\Mpc,0)^{-1}$ has
been estimated to lie between 0.8 (Smoot {\etal} 1992) and 
2.5 (Bardeen {\etal} 1986).
Our pessimistic, middle-of-the-road and optimistic CDM
estimates of $\zvir$ are given in 
Table~\ref{reiontable1}, 
and the
dependence of $\zvir$ on $M_c$ is plotted in 
Figure~\ref{reionfig1}.
This figure  
also shows three alternative models of structure
formation:  CDM with cosmological constant (Efstathiou {\etal} 1992);
tilted CDM (Cen {\etal} 1992) and 
MDM, mixed hot and dark matter
(Shafi
\& Stecker 1984; Schaefer \& Shafi 1992; Davis {\it et al.} 1992;
Klypin {\etal} 1993). For the model with cosmological constant,
we have taken a flat universe with $h=0.5$, $\Omega_0=0.4$ and
$\lambda_0=0.6$. For the tilted model, the power spectrum $P(k)$ is simply
multiplied by a factor $k^{n-1}$, where we have taken $n=0.7$.
For the tilted case, 
\eq{fsEq} still applies. 
For the MDM case,
however, perturbations in the cold component
grow slower than linearly with the scale factor
$(1+z)^{-1}$ and \eq{fsEq} is not valid. 
For the low masses we are considering, we have (Bond \& Szalay 1983) 
$$\sigma_{MDM}(M_c,z) \approx {\sigma_{MDM}(M_c,0)\over (1+z)^{\alpha}},
\quad\hbox{where}$$ 
$$\alpha \equiv {1\over 4}\left[\sqrt{25-24\Omega_{HDM}} - 1\right].$$
Using the parameters from Davis {\etal} (1992), who take
$\Omega=1$ and $\Omega_{HDM}=0.3$, the MDM version of 
\eq{fsEq} becomes 
$$\fs = 
\erfc\left[\left({1+z\over 1+\zvir}\right)^{\alpha}\right]$$
where $\alpha\approx 0.8$ and we redefine $\zvir$ by 
$$(1+\zvir) \approx \left[{\sqrt{2}\,\szmc\over 5.6 \delta_c}\right]^{1/\alpha},$$
with $\szmc$ referring to a pure CDM power spectrum.

For the $\Lambda$ case, perturbations grow approximately linearly until
the universe becomes vacuum dominated at $z \approx \Omega_0^{-1} - 1
= 1.5$, after which their growth slowly grinds to a halt. A numerical
integration of the Friedmann equation and the equation for perturbation 
growth using $h=0.5$, $\Omega_0 = 0.4$ and $\lambda_0 = 0.6$ gives 
$$\sigma_{\Lambda}(M_c,z) \approx 
1.2 {\sigma(M_c,0)\over (1+z)}$$
for $z\gg 3$.

\begin{table}
$$
\begin{tabular}{|l|cccccc|}
\hline
                   & Mixed & Tilted & Lambda & Pess. & Mid. & Opt.\\
\hline
$M_c$&$10^6\Ms$&$10^6\Ms$&$10^6\Ms$&$10^8\Ms$&$10^6\Ms$&$10^5\Ms$\\
Model&MDM&Tilted&Lambda&CDM&CDM&CDM\\
$h$&0.5&0.5&0.5&0.5&0.5&0.8\\
$b$&1&1&1&2&1&0.8\\
$\delta_c$&1.69&1.69&1.69&2.00&1.69&1.44\\
\hline
$\zvir$&2.9&10.1&8.4&4.8&17.2&41.4\\
\hline
\end{tabular}
$$
\vskip-0.3cm
\caption{Galaxy formation assumptions}
\vskip0.3cm
\label{reiontable1}
\end{table}

Since our $\Lambda$-model yields a value of 
$z_{vir}$ very similar to our tilted model, 
we will omit the former from future plots.

\section{Efficiency Parameters}
\label{reionsec3}

In this section, we will discuss the various parameters that give
$\fnet$ when multiplied together. The conclusions are summarized in 
Table~\ref{reiontable2}. 

\begin{table}
$$
\begin{tabular}{|l|rrr|}
\hline
                   & Pess. & Mid. & Opt.\\
\hline
$\fmet$&0.2\%&1\%&25\%\\
$\fesc$&10\%&20\%&50\%\\
$\fuv$&5\%&25\%&50\%\\
$\fion$&10\%&40\%&95\%\\
\hline
$\fnet$&$1\times 10^{-6}$&$2\times 10^{-4}$&$6\times 10^{-2}$\\
\hline
$\fupp$&4&190&24,000\\
\hline
\end{tabular}
$$
\vskip-0.3cm
\caption{Efficiency parameters used}
\vskip0.2cm
\label{reiontable2}
\end{table}

$\fmet$, the fraction of galactic hydrogen that is
burnt into helium during the early life of the galaxy
(within a small fraction of a Hubble after formation),
is essentially the galactic metallicity after the first wave of star
formation. 
Thus it is the product of the fraction of the hydrogen that forms
stars and the average metallicity per star (weighted by mass).
This depends on the stellar mass
function, the galactic star formation rate and the final metallicities
of the high-mass stars.  For our middle-of-the-road estimate, we follow
Miralda-Escud\'e \& Ostriker (1990) in taking $\fmet = 1\%$, half the
solar value.
An upper limit to $\fmet$ is obtained from the extreme scenario where
all the baryons in the galaxy form very massive and short-lived stars
with $M\approx 30\Ms$, whose metallicity could get as high as 25\%
(Woosley \& Weaver 1986). Although perhaps unrealistic, this is not ruled
out by the apparent absence of stars with such metallicities today, since
stars that massive would be expected to collapse into black holes.

In estimating $f_{esc}$, the fraction of the UV photons that despite 
gas and dust manage to escape from the galaxy where they are
created, we follow Miralda-Escud\'e \& Ostriker (1990). 

For $f_{uv}$, the fraction of the released energy that is radiated above
the Lyman limit, we also follow Miralda-Escud\'e \& Ostriker (1990). The
upper limit refers to the extreme $30\Ms$ scenario mentioned above. For
reference, the values of $\fuv$ for stars with various spectra are given in
Table~\ref{reiontable3},
together with some other spectral parameters that will be 
defined and used in the appendix. 
All these parameters involve spectral integrals, and have
been computed numerically.

The parameter $\fion$ is estimated in 
the appendix. 

\begin{table}
$$
\begin{tabular}{|llrrrr|}
\hline
UV source&Spectrum $P(\nu)$&$\fuv$&$\euv$&$\Tpi$&$\uvsigfid$\\
\hline
O3 star&$T=50,000$K Planck&0.57&17.3\,eV&28,300K&2.9\\
O6 star&$T=40,000$K Planck&0.41&16.6\,eV&23,400K&3.4\\
O9 star&$T=30,000$K Planck&0.21&15.9\,eV&18,000K&3.9\\ 
Pop. III star&$T=50,000$K Vacca&0.56&18.4\,eV&36,900K&2.2\\ 
Black hole, QSO&$\alpha=1$ power law&&18.4\,eV&37,400K&1.7\\
?&$\alpha=2$ power law&&17.2\,eV&27,800K&2.7\\
?&$\alpha=0$ power law&&20.9\,eV&56,300K&0.6\\
?&$T=100,000$K Planck&0.89&19.9\,eV&49,000K&1.6\\
\hline
\end{tabular}
$$
\caption{Spectral parameters}
\label{reiontable3}
\end{table}

An altogether different mechanism for converting the baryons in
nonlinear structures into ultraviolet photons is black
hole accretion. If this mechanism is the dominant one, 
\eq{fuppEq} should be replaced by 
$$\fupp \approx
\left({m_pc^2\over 13.6\eV}\right) \fbh\facc\fuv\fesc,$$
where
$$\cases{
\fbh&= mass fraction of nonlinear structures that end up as black
holes,\cr 
\facc&= fraction of rest energy radiated away during accretion
process,\cr  
\fuv&= fraction of energy released as UV photons,\cr
\fesc&= fraction of UV photons that escape from host galaxy.
}$$
There is obviously a huge uncertainty in the factor $\fbh$. 
However, the absence of the factor $0.0073\times 0.76$ compared to
\eq{fuppEq} means that the conversion of matter into 
radiation is so much more efficient 
that the  black hole contribution might be important even
if $\fbh$ is quite small. For instance, $\facc=10\%$ and 
$\fesc=100\%$ gives $\fupp\approx 10^8\fbh\fuv$, which could easily
exceed the optimistic value  $\fupp\approx 24,000$ for the stellar
burning mechanism in 
Table~\ref{reiontable2}. 

\begin{figure}[phbt]
\centerline{\epsfxsize=17cm\epsfbox{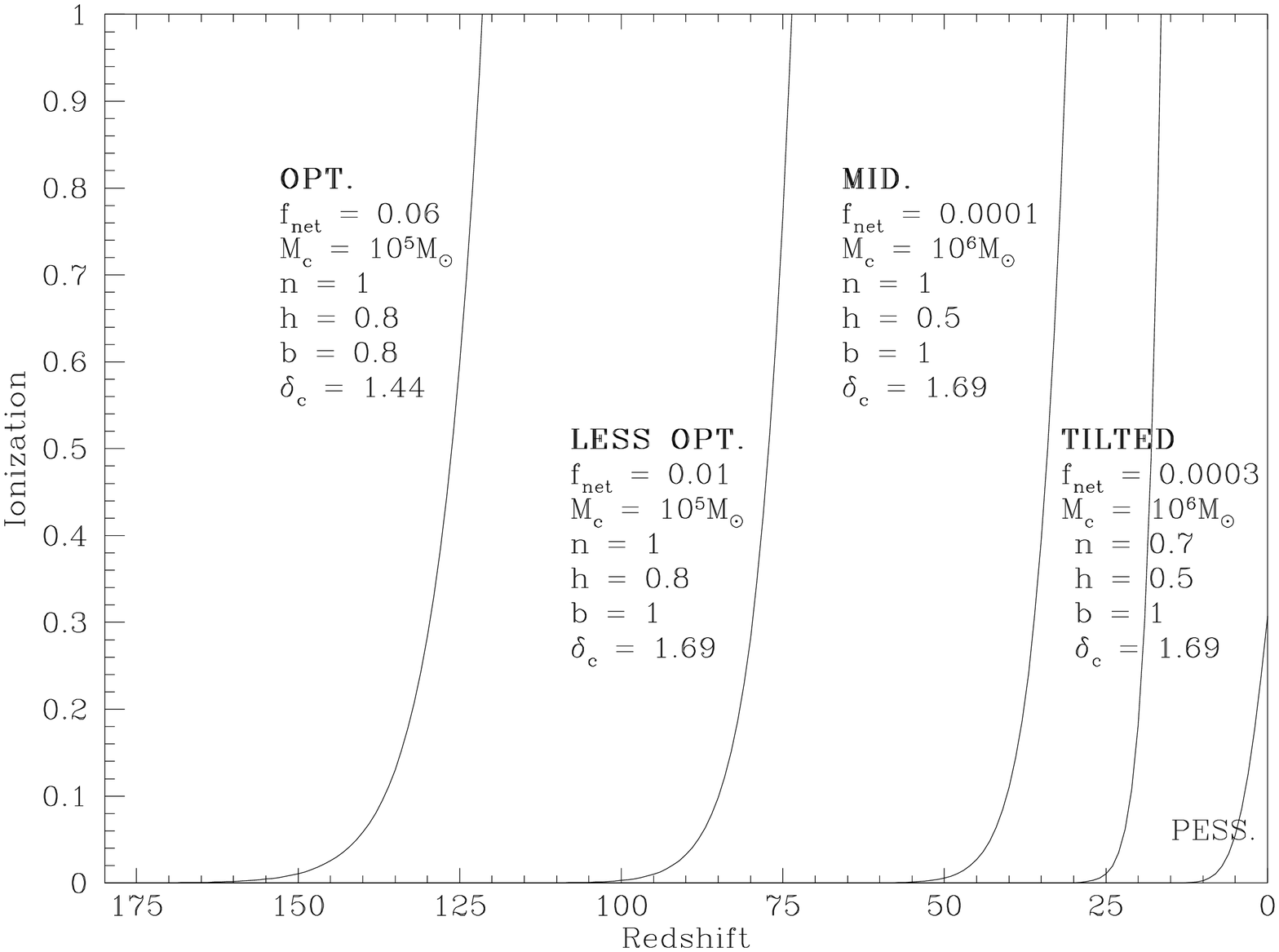}}
\caption{Volume fraction ionized for various scenarios.}
The volume fraction 
of the universe that is in ionized Str\"omgren bubbles
is plotted as a function of redshift for various parameter choices, 
corresponding to $n=1$ CDM (optimistic, less optimistic, middle-of-the-road
and pessimistic cases) and the tilted power spectrum (n=0.7) variant of CDM.
\label{reionfig2}
\end{figure}

\begin{figure}[phbt]
\centerline{\epsfxsize=17cm\epsfbox{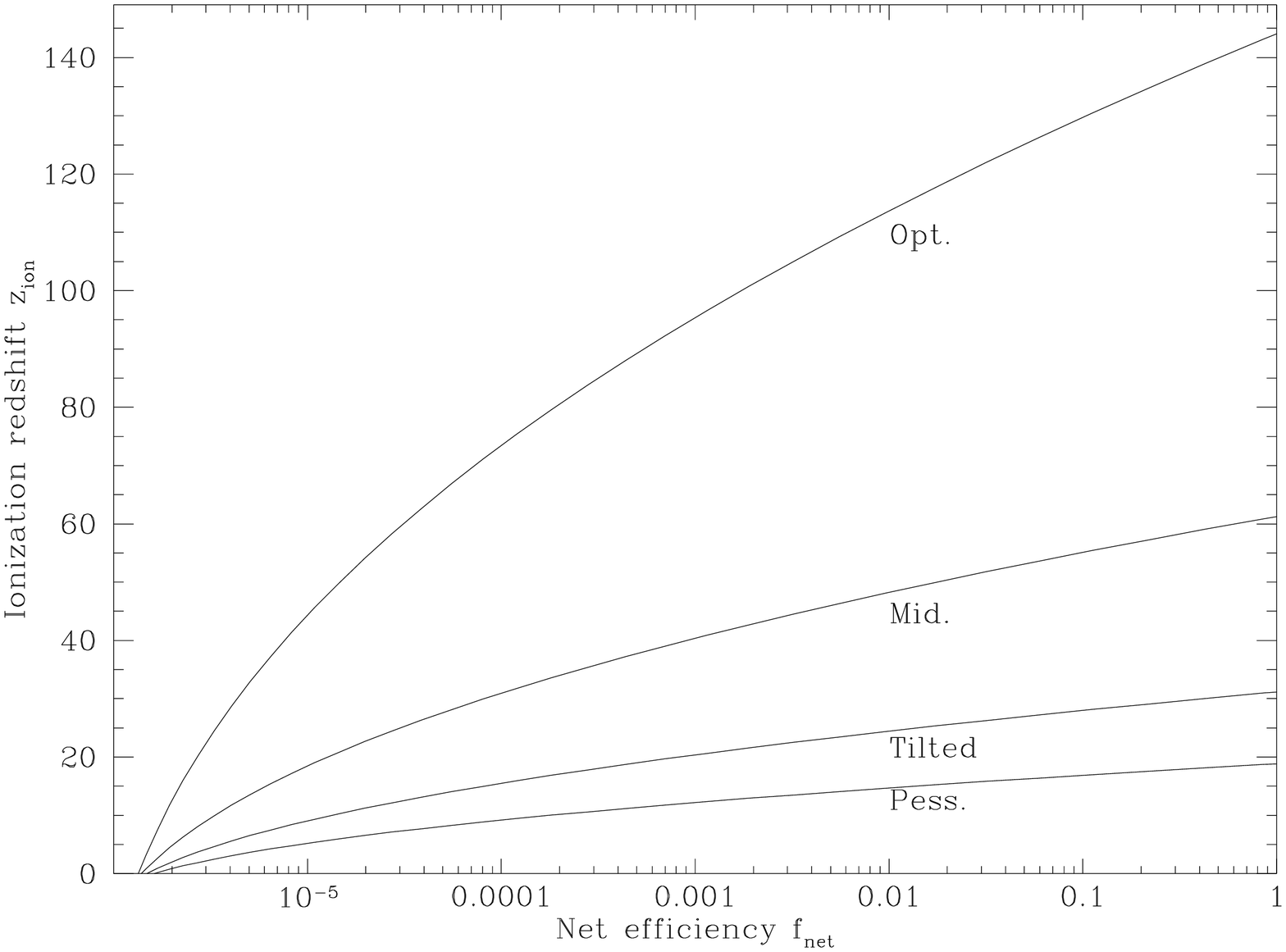}}
\caption{Ionization redshift for various scenarios.}
The redshift at which $x=0.5$ plotted as a function of the net 
efficiency. The four curves correspond to four of the choices of $z_{vir}$
in Table~\ref{reiontable1}: 
$41.4$, $17.2$, $8.4$ and $4.8$ from top to bottom.
\label{reionfig3}
\end{figure}

In 
Figure~\ref{reionfig2},
the ionization fraction 
$\v(z)$ is plotted for various parameter values using
equations\eqnum{FirstFactorEq} and\eqnum{fsEq}.
It is seen that the ionization grows quite abruptly, so that we may speak
of a fairly well-defined {\it ionization redshift}.
Let us define the ionization redshift $\zion$ as the redshift when $\v$
becomes 0.5, {\ie}
\beq{zionEq}
1+\zion = (1+\zvir) \erfc^{-1}\left({1\over 2\fupp\fion}\right).
\eeq
This dependence of $\zion$ of the efficiency is shown in 
Figure~\ref{reionfig3}
for our various galaxy formation scenarios.
It is seen that the ionization redshift is fairly insensitive to the net
efficiency, with the dependence being roughly logarithmic for
$\fnet>0.0001$.

\section{Scattering History}

For a given ionization history $\v(z)$, the Thomson 
opacity out to a redshift $z$, the probability that a CBR
photon is Thomson scattered at least once after $z$, is 
$$\ps(z) = 1 - e^{-\tau(z)},$$ 
where the
optical depth for Thomson scattering is given by
$$\cases{
\tau(z)& = $\tau^*\izz{1+z'\over\sqrt{1+\Omega_0 z'}}\v(z')dz',$\crr
\tau^*& = ${3\Ob\over 8\pi}
\left[1 - \left(1-{\v_{He}\over 4\v}\right)f_{He}\right] 
{H_0c\st\over m_p G} \approx 0.057 h\Ob,$
}$$
where we have taken the mass fraction of helium to be 
$f_{He}\approx 24\%$ and assumed $x_{He} \approx x$,
{\ie} that helium never becomes doubly ionized and that the fraction
that is singly ionized equals the fraction of hydrogen that is
ionized. The latter is a very crude approximation, but makes a
difference of only $6\%$. 
We assume that $\Omega_0=1$ throughout this
paper. 
$\Ob$ denotes the density of the intergalactic medium divided 
by the critical density, and is usually assumed to equal
$\Omega_b$, the corresponding density of baryons. 
the probability that a CBR
photon is Thomson scattered at least once after the standard
recombination epoch at $z\approx 10^3$.

\begin{figure}[phbt]
\centerline{\epsfxsize=17cm\epsfbox{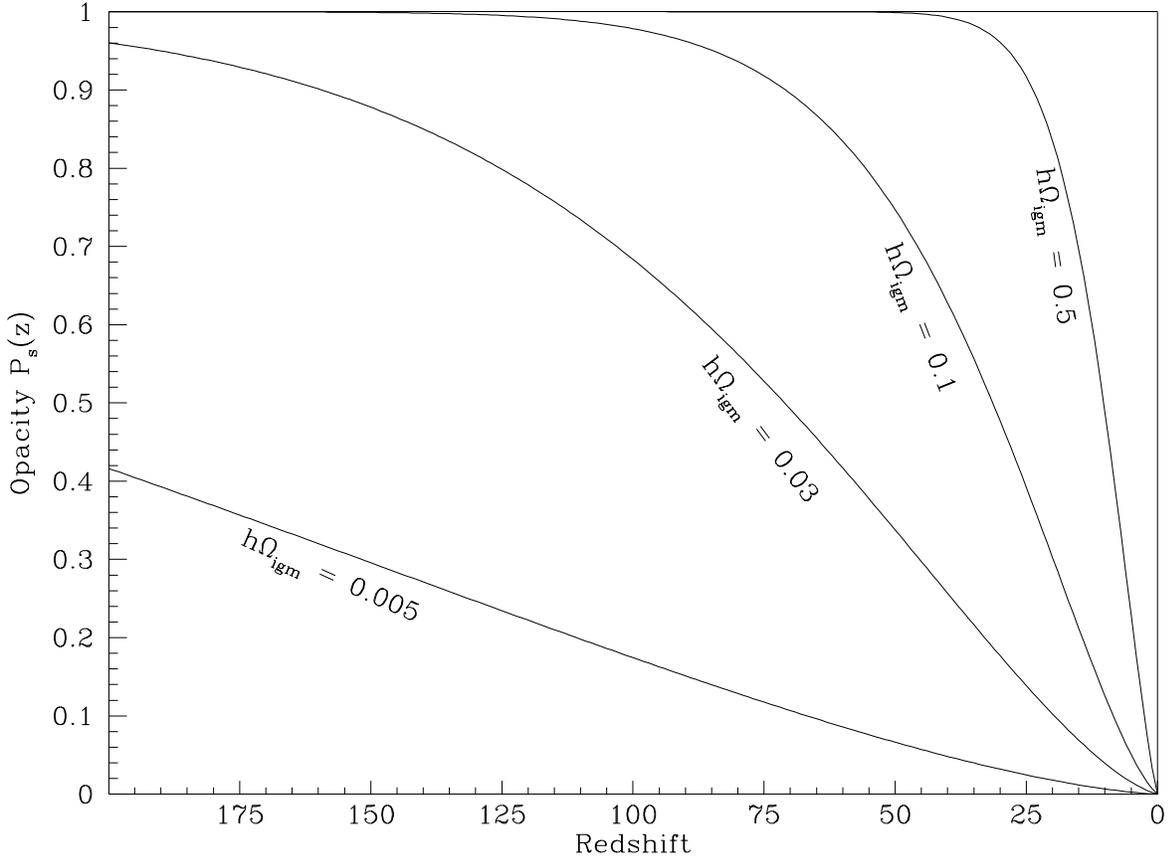}}
\caption{Opacity for completely ionized IGM.}
The Thomson opacity $P_s(z)$, the probability that 
a CBR photon has been scattered at least once after the redshift $z$,
is plotted for four different choices of $h\Ob$ for the case where the
IGM is completely ionized at all times. For more realistic scenarios where
ionization occurs around some redshift $z_{ion}$, the opacity curves
simply 
level out and stay constant for $z\gg z_{ion}$.
\label{reionfig4a}
\end{figure}

\begin{figure}[phbt]
\centerline{\epsfxsize=17cm\epsfbox{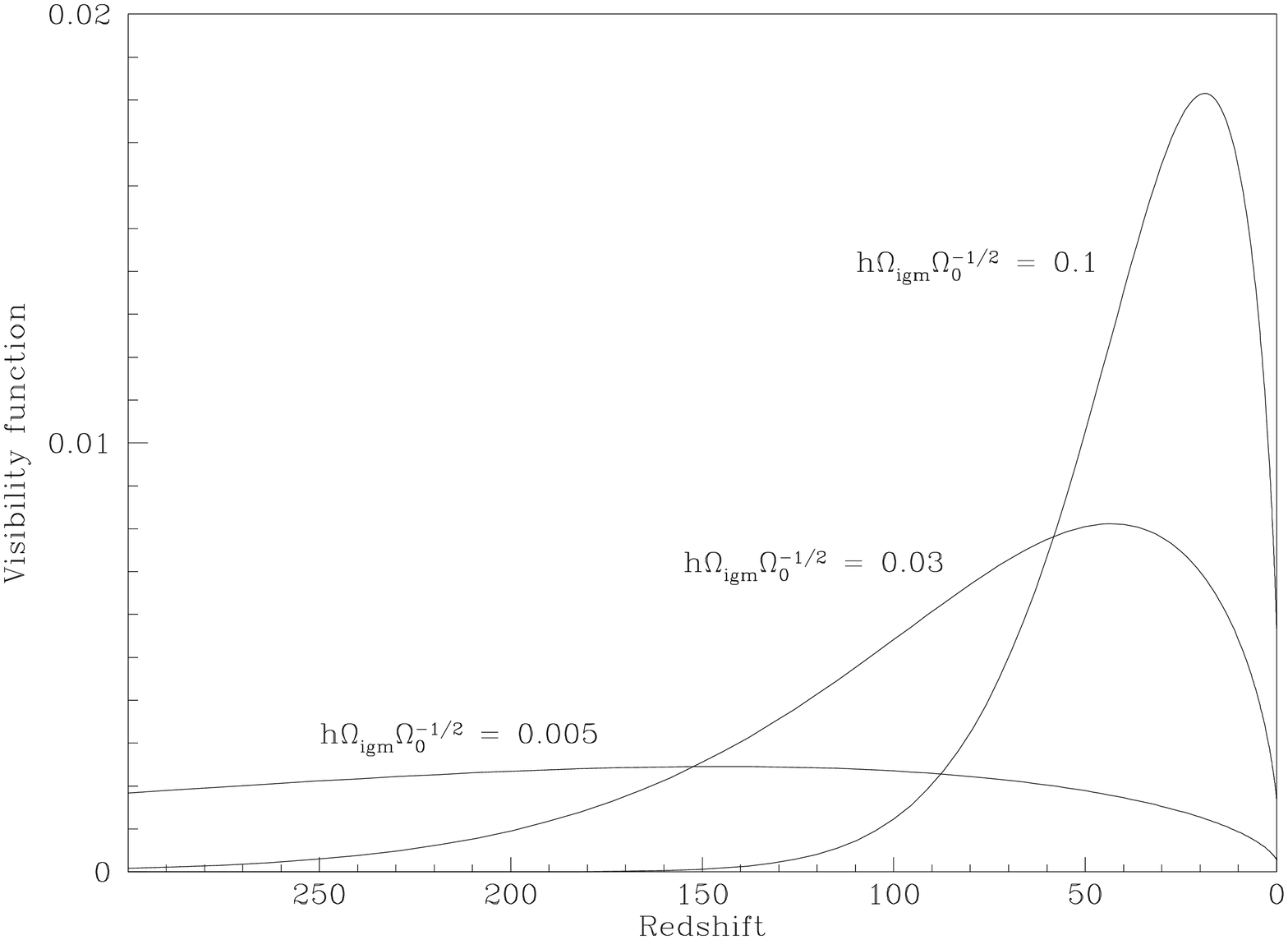}}
\caption{Last-scattering surface for completely ionized IGM.}
The probability distribution for the redshift at which
a CBR photon was last scattered, the so called visibility 
function,  
is plotted for four different choices of $h\Ob$ for the case where the
IGM is completely ionized at all times. For more realistic scenarios where
ionization occurs around some redshift $z_{ion}$, the curves 
are unaffected for $z\ll z_{ion}$, vanish for 
$z_{ion}\ll z\ll 10^3$ and have a second bump around $z\approx 10^3$.\label{reionfig4n}
\label{reionfig4b}
\end{figure}

\begin{figure}[phbt]
\centerline{\epsfxsize=17cm\epsfbox{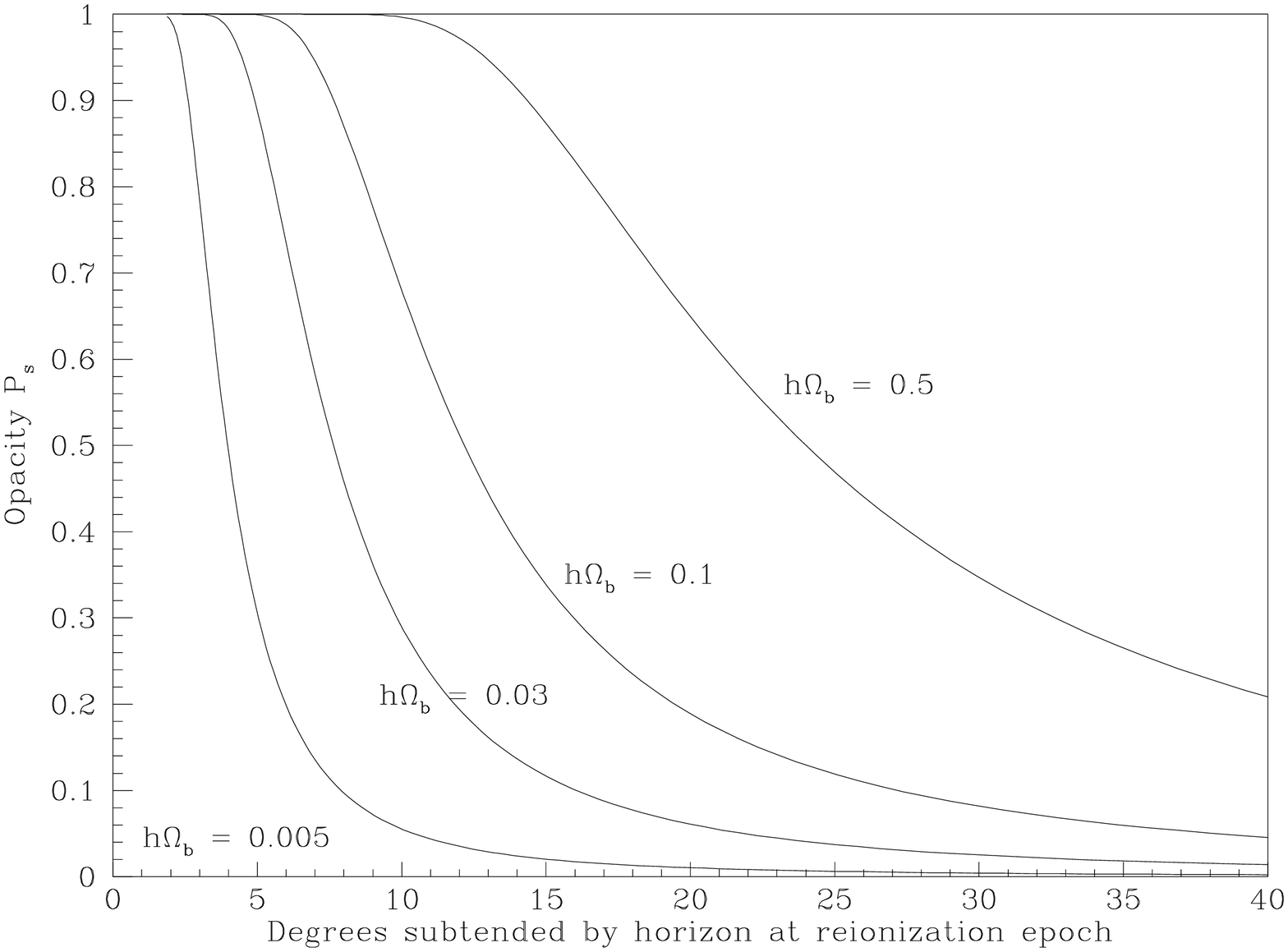}}
\caption{Opacity for completely ionized IGM as function of angle.}
The total Thomson opacity $P_s$, the probability that 
a CBR photon has been scattered at least once since the recombination epoch, 
is plotted as a function of the angle in the sky that the horizon subtended at
the reionization epoch.
This is the largest angular scale on which fluctuations can be suppressed.
\label{reionfig5a}
\end{figure}

\begin{figure}[phbt]
\centerline{\epsfxsize=17cm\epsfbox{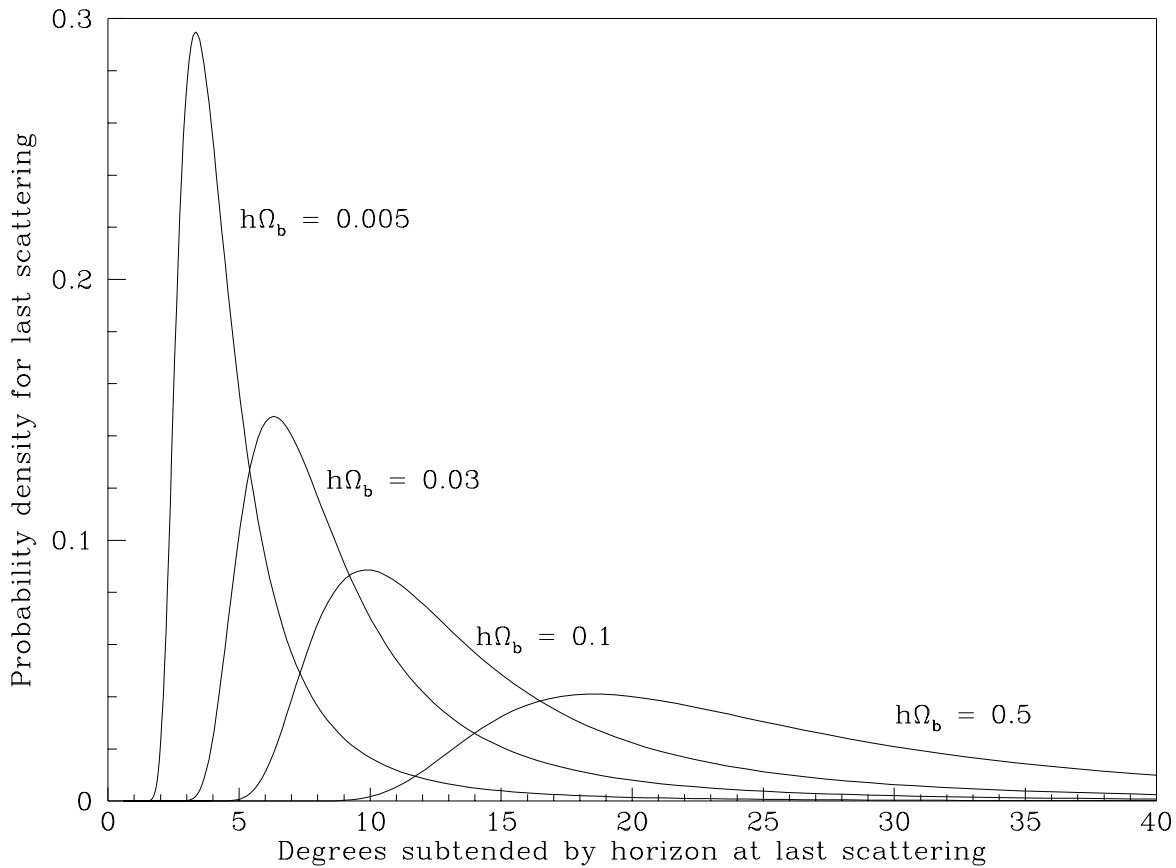}}
\caption{Last-scattering surface for completely ionized IGM as function of angle.}
The probability distribution 
for the angle subtended by the horizon when
a CBR photon was last scattered, the so angular	 visibility 
function,  
is plotted for four different choices of $h\Ob$ for the case where the
IGM is completely ionized at all times. For more realistic scenarios where
ionization occurs around some redshift $z_{ion}$, corresponding to
an angle $\theta_{ion}$, 
the curves 
are unaffected for $\theta\gg \theta_{ion}$, vanish for 
$2^{\circ}\ll \theta\ll \theta_{ion}$ and have a second bump around 
$\theta\approx 2^{\circ}$, the horizon angle at recombination.
\label{reionfig5b}
\end{figure}

The profile of the last scattering surface is given by the so called
visibility function 
$$f_z(z)\equiv {d\ps\over
dz}(z),$$ which is the probability distribution for the redshift 
at which a photon last scattered.
An illuminating special case is that of complete ionization at all
times, {\ie} $\v(z) = 1$, which yields
\beq{PsEq}
\ps(z) =  1-\exp\left(-{2\over
3}\tau^*\left[(1+z)^{3/2} - 1\right]\right) 
\approx 
1-\exp\left[-\left(z\over 92\right)^{3/2}\right]
\eeq
for $z \gg 1$ and $h\Ob=0.03$. 
Hence we see that in order for any
significant fraction of the CBR to have been rescattered
by reionization, the reionization must have occurred quite
early. 
Figures~\ref{reionfig4a} and~\ref{reionfig4b} 
show the opacity and last-scattering
profile for three different choices of $h \Ob$. 
In the optimistic case
$h \Ob = 0.1$, it is seen that even as low an ionization redshift as
$\zion=30$ would give a total opacity $P_s \approx 50\%$.
In 
Figures~\ref{reionfig5a} and~\ref{reionfig5b}, 
we have
replaced $z$ by the angle subtended by the horizon
radius at that redshift, 
$$\theta(z) = 2\arctan\left[{1\over
2\left(\sqrt{1+z}-1\right)}\right],$$
which is the largest angular scale
on which Thomson scattering at $z$ would affect the microwave
background radiation. In 
Figure~\ref{reionfig5b},
we have plotted the angular visibility
function $dP_s/d(-\theta)$ instead of  $dP_s/dz$, so that the curves are
probability distributions over angle instead of redshift.

\begin{figure}[phbt]
\centerline{\epsfxsize=17cm\epsfbox{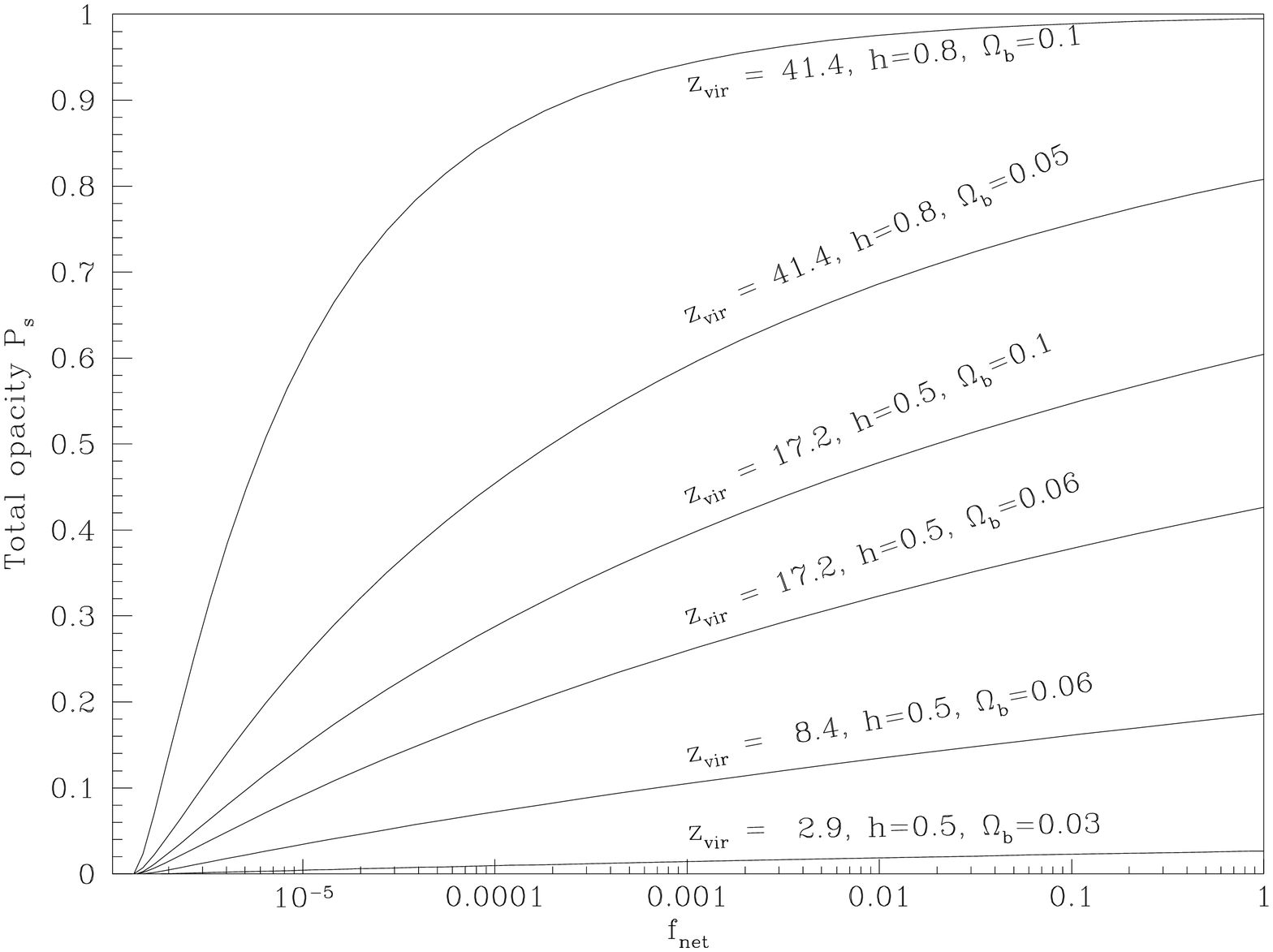}}
\caption{Total opacity for various models.}
The total opacity, the probability that a CBR photon has 
been scattered at least once since the recombination epoch, is plotted for
a variety of models.
\label{reionfig6}
\end{figure}

In the {\it sudden approximation}, the ionization history is a step 
function
$$\v(z) = \theta(\zion - z)$$
for some constant $\zion$, and as was discussed in 
Section~\ref{reionsec3}, 
this
models the actual ionization history fairly well.
In this approximation, the visibility functions are identical to those
in 
Figures~\ref{reionfig4b} and~\ref{reionfig5b} 
for $z<\zion$, but vanish between $\zion$ and
the recombination epoch at $z\approx 10^3$.
Figure~\ref{reionfig6},
which is in a sense the most important 
plot in this paper, shows the 
total opacity $P_s(\zion)$ as a function of $\fnet$ for
a variety of parameter values, as obtained by substituting 
\eq{zionEq} into\eqnum{PsEq}.
As can be seen, the resulting opacity is relatively insensitive to the
poorly known parameter $\fnet$, and depends mainly on the structure
formation model ({\ie} $\zvir$) and the cosmological parameter
$h\Ob$.

Thomson scattering between CBR photons and free electrons affects not
only the spatial but also the spectral properties of the CBR.
It has long been known that hot ionized IGM
causes spectral distortions to the CBR, known as the 
Sunyaev-Zel'dovich
effect. A useful measure of this distortion is the  Comptonization
$y$-parameter (Kompan\'eets 1957; Zel'dovich \& Sunyaev 1969; 
Stebbins \& Silk 1986; Bartlett \& Stebbins 1991)
$$y_c = \int\ktmc n_e\st c\> dt,$$ 
where the integral is to be taken from the reionization epoch to today.  
Let us estimate this integral by making the approximation that the IGM 
is cold and neutral until a redshift $z_{ion}$, at which it suddenly 
becomes ionized, and after which it 
is remains completely ionized with a constant temperature $T$.
Then for $\Omega=1$, $\zion\gg 1$, we obtain 
$$y_c = \ktmc\left({n_{e0}\st c\over H_0}\right)
\int_0^{\zion}\sqrt{1+z}dz \aet{6.4}{-8} h\Ob T_4\> z_{ion}^{3/2},$$
where $T_4\equiv T/10^4\K$ and $n_{e0}$, the electron density today, 
has been computed as before assuming a helium mass fraction of $24\%$
that is singly ionized. 
Substituting the most recent observational constraint 
from the COBE FIRAS experiment,
$y_c < 2.5\times 10^{-5}$ (Mather {\etal} 1994),
into this expression yields
$$\zion < 554T_4^{-2/3} \left({h\Ob\over 0.03}\right)^{-2/3},$$
so all our scenarios are consistent with this spectral constraint.

\section{Discussion}

A detailed discussion of how reionization affects the
microwave background anisotropies would be beyond the scope of this
paper, so we will merely review the main features.
If the microwave background photons are rescattered at a redshift $z$,
then the fluctuations we observe today will be suppressed on angular
scales smaller than the angle subtended by the horizon at that redshift. 
This effect is seen in numerical integrations of the linearized Boltzmann
equation ({\eg} Bond \& Efstathiou 1984; Vittorio \& Silk 1984), and
can be simply understood in purely geometrical terms. 
Suppose we detect a microwave photon arriving from some direction in
space. Where was it just after recombination? In the absence of
reionization, it would have been precisely where it appears to be coming
from, say $3000$ Mpc away. If the IGM was reionized, however, the photon
might have originated somewhere else, scattered off of a free electron and
then started propagating towards us, so at recombination it might even
have been right here. Thus to obtain the observed anisotropy, we have to
convolve the anisotropies at last scattering with a window function
that incorporates this smoothing effect. Typical widths for the window
function appropriate to the last scattering surface range from a few
arc-minutes with standard recombination to the value of a few degrees
that we have derived here for early reionization models. 

In addition to
this suppression on sub-horizon scales, new fluctuations will be
generated by the first order Doppler effect and by the Vishniac
effect. The latter dominates on small
angular scales and is not included in the linearized Boltzmann
treatment because it is a second order effect.
The current upper limit on CBR fluctuations on the 1 arcminute scale
of $\Delta T/T < 9\tento{-6}$ (Subrahmanyan {\etal} 1993) provides
an interesting constraint on reionization histories through the
Vishniac effect.
In fact, according to the original calculations (Vishniac 1987), 
this would rule out most of the reionization histories in this paper.
However, a more careful treatment (Hu {\etal} 1994) predicts a Vishniac
effect a factor of five smaller on this angular scale, so all
reionization histories in this paper are still permitted.

The COBE DMR detection of $\Delta T/T$ has provided a normalization for
predicting CBR anisotropies on degree scales. Several experiments are
underway to measure such anisotropies, and early results that report
possible detections have recently become available from experiments at
the South Pole (Meinhold \& Lubin 1991; Shuster {\etal} 1993) and at
balloon altitudes (Devlin {\etal} 1992; Meinhold {\etal} 1993;
Shuster {\etal} 1993). There is some reason to believe that these detected
signals are contaminated by galactic emission. Were this the case, the
inferred CBR upper limits to fluctuations on degree scales might be
lower than those predicted from COBE extrapolations that adopt the
scale-invariant power spectrum that is consistent with the DMR result and
is generally believed to be the most appropriate choice on large scales
from theoretical considerations 
({\eg} Gorski {\etal} 1993; Kashlinsky 1992). 
In the absence of such contaminations, the
detected fluctuations in at least some degree-scale experiments are,
however, consistent with the COBE extrapolation ({\eg} Jubas \&
Dodelson 1993). The variation from field to field, repeated on
degree scales, also may argue either for galactic contamination or else for
unknown experimental systematics, or even non-Gaussian fluctuations.
The results of other recent experiments 
such as ARGO (de Bernardis {\etal} 1993),
PYTHON (Dragovan {\etal} 1993) and MSAM (Cheng {\etal} 1993) have reinforced the 
impression that the experimental data is not entirely self-consistent, and 
that some form of systematic errors may be important.

The controversy over the interpretation of the degree-scale CBR
fluctuations makes our reanalysis of the last scattering surface
particularly timely. We have found that canonical dark matter, tailored
to provide the 10 degree CBR fluctuations detected by the COBE DMR
experiment,
results in sufficiently early reionization (before $z \approx 50$)
over a fairly wide range of parameter space, to smooth out
primordial degree-scale  fluctuations. Our middle-of-the-road model
produces suppression by roughly a factor of two; it is difficult,
although not impossible, to obtain a much larger suppression.
This smoothing, because it is of order unity in scattering optical
depth, is necessarily inhomogeneous. We predict the presence of
regions with large fluctuations and many ``hot spots" and ``cold spots", 
corresponding to ``holes" in the last-scattering
surface, as well as regions with little small-scale power where the last 
scattering is more efficient. The
detailed structure of the CBR sky in models with reionization
will be left for future studies.
Here we
simply conclude by emphasizing that anomalously low values of $\Delta
T/T$ over degree scales are a natural corollary of reionization at
high redshift.

\appendix
\section{Appendix: The Efficiency Parameter $\fion$}

In this appendix, we will discuss the parameter $\fion$, and see that
it rarely drops below $30\%$. 
We will first discuss the thermal
evolution of intergalactic hydrogen exposed to a strong UV flux, and then
use the results to write down a differential equation for the
volume fraction of the universe that is ionized, subject to 
point sources of UV radiation that switch on at different times. 
We will see that photoionization is so
efficient within the ionized regions of the IGM that
quite a simple equation can be given for the expansion of 
the ionized regions. 

The evolution of IGM exposed to ionizing radiation has been discussed by many
authors. Important early work includes that of Arons \& McCray (1970),
Bergeron \& Salpeter (1970) and 
Arons \& Wingert (1972). 
The main novelty of the treatment that follows is that 
whereas previous treatments focus on late ($z<5$) epochs, when 
various simplifying approximations 
can be made because the recombination and Compton rates are low,
we are mainly interested in the case $50<z<150$.
We show that IGM exposed to a strong UV flux rapidly approaches
a quasistatic equilibrium state, where it is almost fully ionized and
the temperature is such that 
photoionization heating exactly balances Compton cooling.
This simplifies the calculations dramatically, since the entire thermal 
history of
the IGM can be summarized by a single function $\v(z)$, the volume 
fraction that is ionized. 
Thus a fraction $\v(z)$ is ionized and hot (with a temperature 
that depends only on $z$, not on when it became ionized), 
and a fraction $1-\v(z)$ is neutral and cold.

In the first section, we justify this approximation.
In the second section, we derive a differential equation for the
time-evolution of $\v$ as well as a useful analytic estimate
of $\fion$.

\subsection{Intergalactic Str\"omgren Spheres}

Let $\i$ denote the ionization fraction in a small, 
homogeneous volume of intergalactic hydrogen, {\ie}
$\i\equiv n_{HII}/(n_{HI}+n_{HII})$. 
($\i$ is not to be confused with $\v$, the volume fraction in
ionized bubbles.)
When this IGM is  
at temperature $T$, exposed to a
density of $\eta$ UV photons per proton, the
ionization fraction $\i$ evolves as follows: 
\beq{aIonizationEq}
{d\i\over d(-z)} = 
{1+z\over\soz}
\left[\lpi(1-\i) + \lci\i(1-\i) - \lrec^{(1)}\i^2\right],
\eeq
where $H_0^{-1}(1+z)^{-3}$ times the rates per baryon for
photoionization, collisional ionization and recombination are given by
\beq{aRateEq}
\cases{
\lpi \aet {1.04}{12} \left[h\Ob\uvsigfid\right]\eta,&\crr
\lci \aet{2.03}{4} h\Ob T_4^{1/2} e^{-15.8/T_4,}&\crr
\lrec^{(1)} \approx 0.717 h\Ob 
T_4^{-1/2}\left[1.808-0.5\ln T_4 + 0.187 T_4^{1/3}\right],&
}
\eeq
and $T_4\equiv T/10^4\K$. $\uvsigfid$ is the thermally averaged 
photoionization cross section in units of $10^{-18}\cm^2$, and has been
computed in 
Table~\ref{reiontable3} 
for various spectra 
using the differential cross section
from Osterbrock (1974). The collisional ionization rate is from Black (1981).
The recombination rate is the total rate to all hydrogenic levels
(Seaton 1959). 

Below we will see that in the ionized Str\"omgren bubbles that will
appear around the galaxies or quasars, the photoionization
rate is so much greater than the other rates that to a good
approximation, \eq{aIonizationEq} can be replaced by the
following simple model for the IGM:

$\bullet$ It is completely ionized ($\i=1$).

$\bullet$ When a neutral hydrogen atom is formed
through recombination, it is instantly photoionized again.

Thus the only unknown parameter is the IGM temperature, which
determines the recombination rate, which in turn equals the
photoionization rate and thus determines the rate of heating.

Let us investigate when this model is valid.
Near the perimeter of an ionized Str\"omgren sphere of radius $r$
surrounding a galaxy, the number of UV photons per proton is roughly 
$$\eta = {S_{uv}\over
4\pi r^2 c n},$$ 
where $S_{uv}$ is the rate at which UV photons leave the galaxy. 
For an O5 star, the photon flux above the Lyman limit is 
approximately $3.1\times 10^{49}\s^{-1}$ (Spitzer 1968), so if each 
$N$ solar masses of baryons in a galaxy leads to production of a UV
flux equivalent to that of an O5 star, then 
\beq{etaEq}
\eta \geq 0.77 {\fesc M_6\over h^2 r_1^2 N (1+z)^3},
\eeq
inside the sphere, where $r_1\equiv r/1\Mpc$ and
$M_6\equiv M/10^6\Ms$.
When a fraction $f_s$ of all matter has formed galaxies of a typical
total (baryonic and dark) mass $M$, then in the absence of strong
clustering, the typical separation between two galaxies is $$R =
\left({M\over\fs\rho}\right)^{1/3}\approx  \left({15\kpc\over 1+z}\right)
\left({M_6\over h^2\fs}\right)^{1/3},$$
where $M_6\equiv M/10^6\Ms$. Thus $r$ continues to increase until 
$r\approx R$, and spheres from neighboring galaxies begin to overlap.
We are interested in the regime where $z<150$. Substituting this and 
\eq{etaEq} into\eqnum{aRateEq}, we see that
$\lpi\gg\lci$ and $\lpi\gg\lrec$ for any reasonable parameter values.
Hence we can neglect collisional ionization in 
\eq{aIonizationEq}. Since $\lpi\gg 1$,
the photoionization timescale is much shorter than the Hubble
timescale, so \eq{aIonizationEq} will quickly approach a
quasistatic equilibrium solution where the recombination rate equals the
photoionization rate, {\ie}
$$\i\approx 1-{\lrec\over\lpi}\approx 1.$$
In conclusion, the simple $\i=1$ model is valid for all parameter
values in our regime of interest.

When a hydrogen atom gets ionized, the photoelectron acquires 
an average
kinetic energy of ${3\over 2} k\Tpi$, where $\Tpi$ is defined by
${3\over 2}k\Tpi = \euv - 13.6\eV$, and $\euv$ is the average energy of
the ionizing photons 
(see Table~\ref{reiontable3}).

Since the timescale for Coulomb
collisions is much shorter than any other timescales involved,
the electrons and protons rapidly thermalize, and we can always assume
that their velocity distribution is Maxwellian, corresponding to some 
well-defined temperature $T$.
Thus shortly after the 
hydrogen gets photoionized, after the electrons have 
transferred half of their energy to the protons, the plasma temperature is 
$T = {1\over 2} \Tpi$.

The net effect of a recombination and subsequent photoionization is to
remove the kinetic energy of the captured electron,
say ${3\over 2} kT\recfac(T)$, from 
the gas and replace it with
${3\over 2} k\Tpi$,
the kinetic energy 
of the new photoelectron.
Since the recombination cross section is approximately proportional
to $v^{-2}$, slower electrons are
more likely to get captured. Hence the mean energy of the captured electrons 
is slightly lower than ${3\over 2}kT$, {\ie} $\recfac(T)$ is slightly less
than unity (Osterbrock 1974). 
We compute $\recfac(T)$ using Seaton (1959). 
The complication that $\recfac(T)\neq 1$ turns out to 
be of only marginal importance: 
$\recfac(10^4\K)\approx 0.8$, which only 
raises the equilibrium temperatures calculated below by a few percent.

The higher the recombination
rate, the faster this effect will tend to push the temperature 
up towards
$\Tpi$. The two dominant cooling effects are Compton drag against the
microwave background photons and cooling due to the adiabatic expansion of
the universe. Line cooling from collisional excitations 
can be neglected, since the neutral fraction $1-\i \approx 0$.
 Combining these effects, we obtain the evolution equation
for the IGM inside of a Str\"omgren bubble:
\beq{aTeq}
{dT\over d(-z)} = -{2\over 1+z}T + 
{1+z\over\sqrt{1+\Oz z}}\left[\lcomp(T_{cbr}-T) +
{1\over 2}\lrec(T)[T_{cbr}-\eta_{rec}(T)T]\right]
\eeq
where
$$\lcomp = {4\pi^2\over 45}
\left({k T_{cbr}\over\hbar c}\right)^4
{\hbar\st\over H_0 m_e}(1+z)^{-3}
\approx 0.00418 h^{-1}(1+z)$$
is $(1+z)^{-3}$ times the Compton cooling rate per Hubble time
and $T_{cbr} = T_{cbr,0}(1+z)$.
We have taken $T_{cbr,0}\approx 2.726\K$ (Mather {\etal} 1994).
The factor of ${1\over 2}$ in front of the $\lrec$ term is due to the
fact that the photoelectrons share their acquired energy with the protons.
The average energy of
the ionizing photons is given by the spectrum $P(\nu)$ as
$\euv = h\expec{\nu}$,
where 
$$\expec{\nu} = 
{\izi P(\nu)\sigma(\nu) d\nu\over
 \izi \nu^{-1}P(\nu)\sigma(\nu) d\nu}.$$
Here the photoionization cross section
$\sigma(\nu)$ is given by
Osterbrock (1974). Note that, in contrast to certain nebula
calculations where all photons get absorbed sooner or later, the
spectrum should be weighted by the photoionization cross section. This
is because most photons never get absorbed in the Str\"omgren  regions
(only in the transition layer), and all that is relevant is the energy
distribution of those photons that do. $P(\nu)$ is the energy
distribution (W/Hz), not the number distribution which is proportional
to  $P(\nu)/\nu$.

The spectral parameters $\euv$ and $\Tpi$ are given in 
Table~\ref{reiontable3}
for some selected spectra. A Planck spectrum 
$P(\nu)\propto\nu^3/\left(e^{h\nu/kT}-1\right)$
gives quite a good prediction of $T^*$ for stars with surface temperatures
below $30,000\K$. For very hot stars, more realistic spectra (Vacca, 1993)
have a sharp break at the Lyman limit, and fall off much slower above it,
thus giving higher values of  $T^*$. As seen in 
Table~\ref{reiontable3},
an
extremely metal poor star of surface temperature $50,000\K$ gives roughly the
same $T^*$ as QSO radiation. 
The only stars that are likely to be relevant to
early photoionization scenarios are hot and short-lived ones,
since the universe is only about $10^7$ years
old at $z = 100$, and fainter stars would be unable to inject enough
energy in so short a time. Conceivably, less massive stars could play
a dominant role later on, thus lowering $T^*$. However, since they
radiate such a small fraction of their energy above the Lyman limit,
very large numbers would be needed, which could be difficult to
reconcile with the absence of observations of Population III stars
today. If black holes are the dominant UV source, the stellar spectra
of 
Table~\ref{reiontable3}
are obviously irrelevant. A power law spectrum
$P(\nu)\propto \nu^{-\alpha}$ with $\alpha=1$ fits observed QSO
spectra rather well in the vicinity of the Lyman limit (Cheney \&
Rowan-Robinson 1981; O'Brien {\etal} 1988), and is also consistent
with the standard model for black hole accretion. 

\begin{figure}[phbt]
\centerline{\epsfxsize=17cm\epsfbox{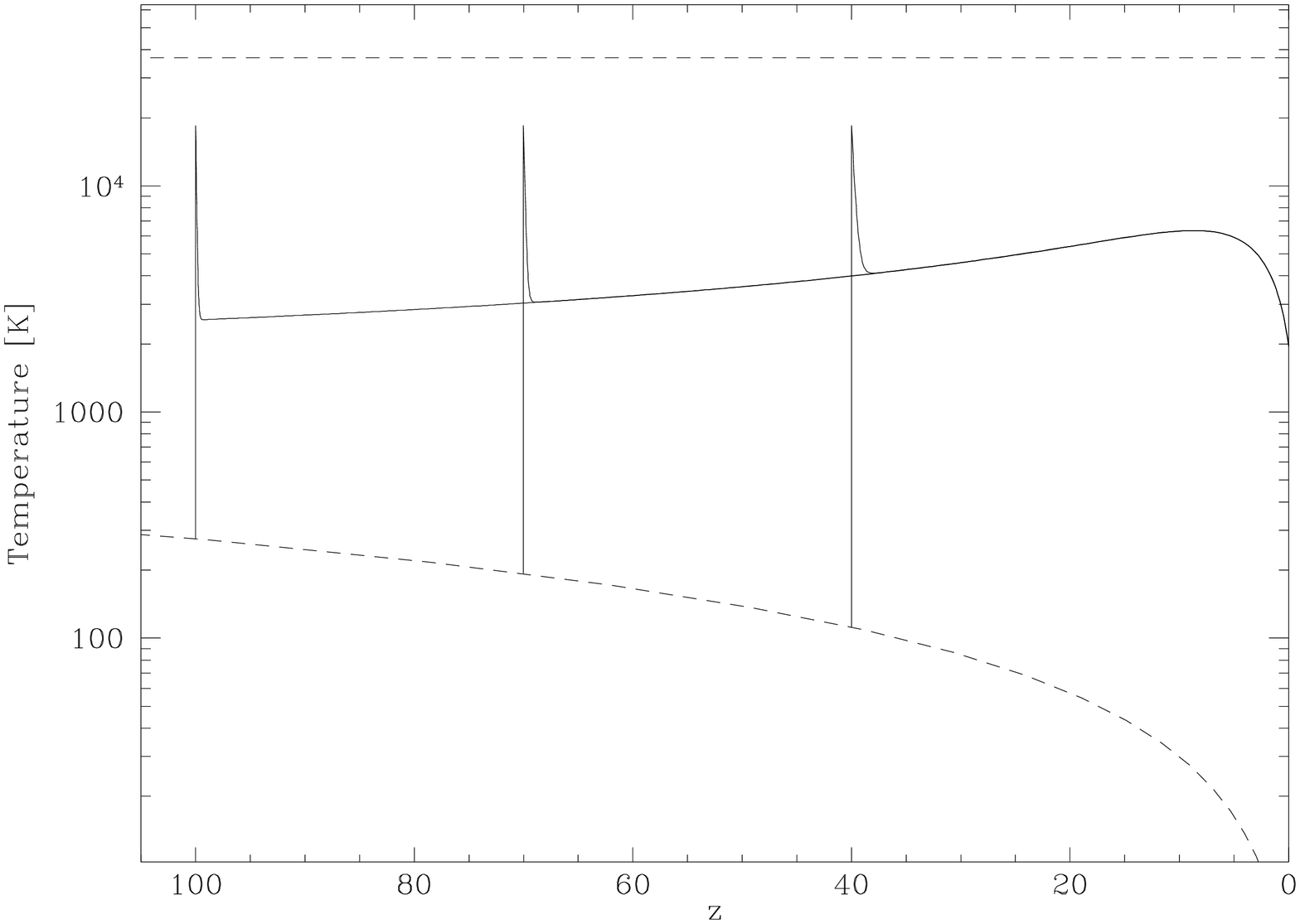}}
\caption{Temperature evolution in intergalactic Str\"omgren bubbles.}
The temperature evolution is plotted 
for IGM exposed to a UV flux 
strong enough to keep it completely photoionized. In this example,
$h=0.5$, $\Ob=0.06$, and $T^* = 36,900$.  
The upper dashed line is $T^*$, the temperature corresponding to 
the average energy of the released photoelectrons,
towards which the
plasma is driven by recombinations followed by new photoionizations. 
The lower dashed line is the temperature of 
the CBR photons, towards which the plasma is driven Compton cooling. 
The three solid curves from left to right correspond to three different
redshifts for becoming part of a Str\"omgren bubble.
The first time the hydrogen becomes ionized, 
its temperature rises impulsively
to $T^*/2$. After this, Compton cooling 
rapidly pushes the temperature down to
a quasi-equilibrium level, where the Compton cooling rate equals the
recombination heating rate.
\label{reionfig7}
\end{figure}

Numerical solutions to 
\eq{aTeq} are shown in 
Figure~\ref{reionfig7},
and it is
seen that the temperature evolution separates into three distinct phases.
In the first phase, 
the IGM is outside of the Str\"omgren regions, 
unexposed to UV radiation, and 
remains cold and neutral.
In the second phase, the IGM suddenly becomes ionized, and its temperature
instantly rises to ${1\over 2}\Tpi$. After this, Compton cooling rapidly 
reduces the temperature to a
quasi-equilibrium value of a few thousand K.
After this, in the third phase, $T$ changes only quite slowly,
and is approximately given by setting the expression
in square brackets in \eq{aTeq} equal to zero.
This quasi-equilibrium temperature is typically many times lower
than $T^*$, since Compton cooling is so efficient at the high 
redshifts involved.

\subsection{The Expansion of Str\"omgren Regions}

This rapid approach to quasi-equilibrium, where the IGM 
``loses its memory" of how long ago it became part of a 
Str\"omgren region,
enables us to construct a very simple model for
the ionization history of the universe.
At redshift $z$, a volume fraction $\v(z)$ of the universe is
completely ionized and typically has a temperature of a few thousand K.
The ionized part need not consist of non-overlapping
spheres; it can have any topology whatsoever. The remainder is cold and
neutral.

Between the ionized and neutral regions is a
relatively thin transition layer, where 
the IGM becomes photoionized and its temperature 
adjusts to the quasistatic value as in 
Figure~\ref{reionfig7}.
As this IGM becomes part of the hot and ionized
volume, the transition layer moves, and $\v(z)$ increases\footnote
{We are tacitly assuming that the UV luminosity of the galaxy that 
creates each Str\"omgren sphere never decreases.
Although obviously untrue, this is in fact an excellent approximation,
since these early dwarf-galaxies correspond to perturbations 
far out in the Gaussian tail.
Since $\fs(z)$ grows so dramatically as the redshift decreases and we
move from five sigma to four sigma to three sigma, {\it etc.}, almost 
all galaxies in existence at a given redshift 
are in fact very young, so that 
older ones that have begun to dim can be safely neglected.
}.

As long as $\v<1$, all UV photons produced are absorbed instantly to a
good approximation. Thus the rate at which UV photons are released
is the sum of the rate at which they are used to counterbalance
recombinations inside the hot bubbles and the rate at which they are
used to break new ground, to increase $\v$. Thus
$$\fupp{d\fs\over dt} = \alpha^{(2)}(T) n\v + {d\v\over dt},$$
where $\alpha^{(2)}(T)$ is the total recombination rate 
to all hydrogenic levels except the ground state\footnote
{A reionization directly to the ground state produces a UV photon that 
usually propagates uninterrupted through the 
highly ionized Str\"omgren region, and then ionizes another atom 
in the transition layer
between the
expanding Str\"omgren region and its cold and neutral surrounding.
Thus 
recombinations directly to the ground state 
were included in the above calculation of the quasi-equilibrium
temperature of the Str\"omgren bubbles, since 
the resulting UV photons could be considered lost from 
the latter.
Here, on the contrary, recombinations directly to the ground state
should not be included, since the UV photons they produce are not wasted
from an energetics point of view.}.
Changing the independent
variable to redshift and using \eq{fsEq}, 
we find that this becomes
\beq{ExpansionEq}
{d\v\over d(-z)} + \lrec^{(2)}{1+z\over\sqrt{1+\Omega_0 z}}\v = 
{2\over\sqrt\pi}\left({\fupp\over 1+\zvir}\right)
\exp\left[-\opzopzv^2\right].
\eeq
Here 
$$\lrec^{(2)} \approx 0.717 h\Ob 
T_4^{-1/2}\left[1.04-0.5\ln T_4 + 0.19 T_4^{1/2}\right]$$
is $H_0^{-1}(1+z)^{-3}$ times 
the total recombination rate per 
baryon to all hydrogenic levels except the ground
state. The fit is to the data of Spitzer (1968) and is accurate to
within $2\%$ for $30\K<T<64,000\K$.
$\lrec^{(2)}$ is to be evaluated at the quasi-equilibrium temperature 
$T(z)$ discussed above.

Using the values in 
Table~\ref{reiontable3}
for the pessimistic, middle-of-the-road and
optimistic estimates, the parameter $\fupp$ equals 
roughly 4, 190 and 24,000, 
respectively.

In the absence of photon waste through recombination, equation
\eq{ExpansionEq} would have the solution 
$\v^*(z) = \fupp\fs(z),$
so the ionization efficiency is
$$\fion(z) = \v(z)/\v^*(z).$$ 
Since \eq{ExpansionEq} is linear in $\v$ and the initial
data is $\v=0$ at some redshift, it is readily seen that the solution
$\v(z)$ is proportional to $\fupp$, the constant in front of the
source term. Combining these last two observations, we see that 
$\fion$ is independent of $\fupp$ and hence independent of the
poorly known parameters $\fmet$, $\fuv$ and $\fesc$.

\begin{figure}[phbt]
\centerline{\epsfxsize=17cm\epsfbox{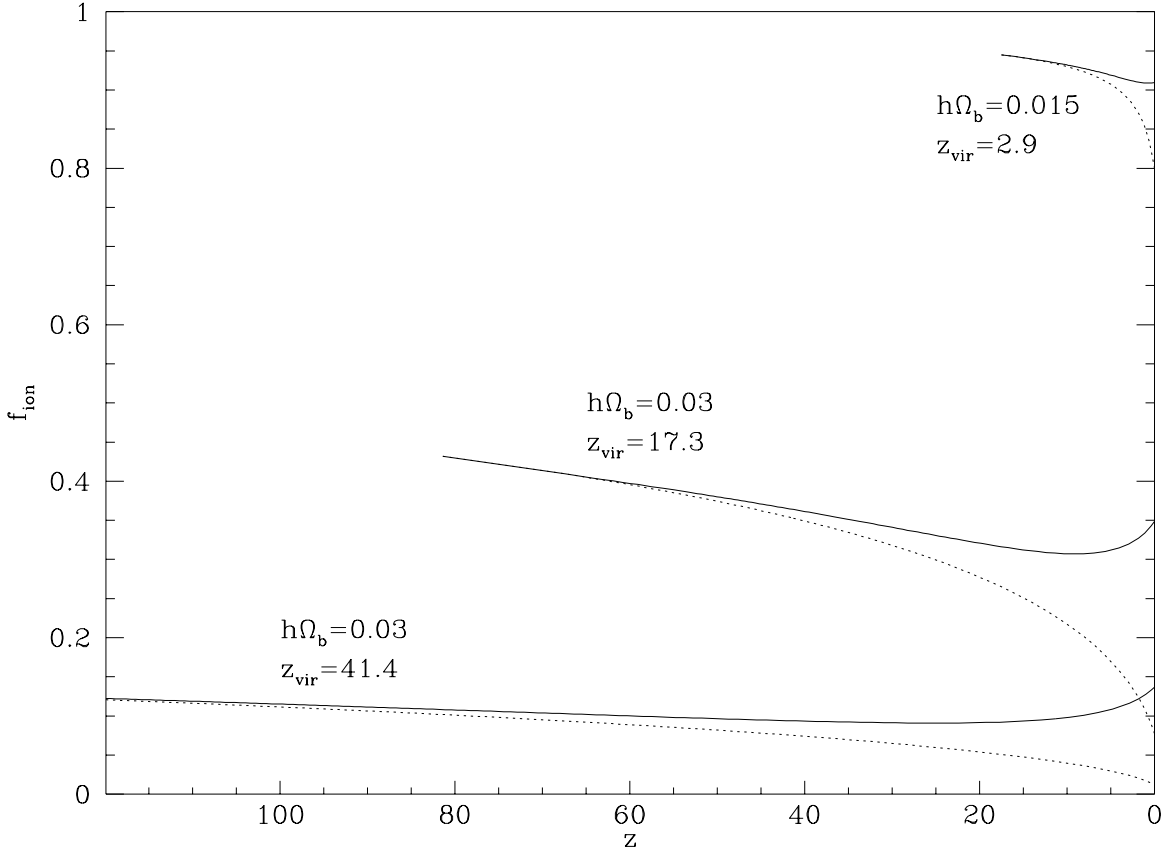}}
\caption{Ionization efficiencies for various scenarios.}
The ionization efficiency, the 
fraction of the UV photons that produce a net 
ionization, is plotted for three different parameter combinations.
In all cases, $T^*=36,900\K$, the value appropriate for the radiation 
from the population 3 star in Table~\ref{reiontable3}.
The solid lines are the exact results from numerical integration of
\eq{ExpansionEq}. The dotted lines are the analytic fits, which are seen
to agree well in the redshift range of interest, which is typically
$z$ twice or three times $z_{vir}$.
\label{reionfig8}
\end{figure}

Plots of $\fion(z)$ from numerical solutions 
of \eq{ExpansionEq} are shown in 
Figure~\ref{reionfig8}
for various
parameter values, and it is seen that the dependence on $z$ is generally
quite weak. Let us make use of this fact by substituting the 
{\it Ansatz} 
$\v(z) = \fion(z)\fupp\fs(z)$ into \eq{ExpansionEq}, 
and setting $\fion'(z)\approx 0$. 
Using \eq{fsEq} and an asymptotic
approximation for the error function, we obtain
$$\fion(z) \approx
{1\over 1 + 0.48{\lrec^{(2)}(1+\zvir)^2/\soz}},$$
independent of $\fupp$, which agrees to within $10\%$ with the numerical
solutions for all reasonable parameter values.
This expression highlights 
the connection between $\fion$ and the thermal evolution
of the Str\"omgren bubbles: 
Essentially, the higher the quasi-static temperature,
the lower the recombination rate $\lrec^{(2)}$, and
the higher $\fion$ becomes.

The value of $\fion$ relevant to computing the ionization redshift is 
obviously that where $z=\zion$. As we have seen, 
$\zion$ typically lies between $2\zvir$ and $3\zvir$. 
Substituting $T\approx 2,500\K$ into the expression for $\lrec$ and
taking $\Omega_0 \approx 1$ and $z=\zion\approx 2.5\zvir$, the above reduces to  
$$\fion \approx
{1\over 1 + 0.8 h\Ob(1+\zvir)^{3/2}},$$
so we see that $\fion$ will be of order unity
unless $\zvir\gg 15$ or $h\Ob\gg 0.03$.

\bigskip
The authors would like to thank Prof. A. Blanchard for
illuminating discussions on the subject of the paper and 
W. Hu, A. Reisenegger, D. Schlegel, D. Scott and our referee 
for many useful comments.
This research has been supported in part by a grant from the NSF.

%%%%%%%%%%%%%%%%%%%%%% REFERENCES: %%%%%%%%%%%%%%%%%%%%%%%%%

\section{REFERENCES}

\rf Arons. J. \& McCray, R. 1970; Ap. Letters;5;123

\rf Arons, J. \& Wingert, D. 1972; Ap. J.;177;1

\rf Bardeen, J. M., Bond, J. R., Kaiser, N. \& Szalay, A. S.
1986;Ap. J.;304;15

\rf Bartlett, J. \& Stebbins, A. 1991;Ap. J.;371;8

\rf Bergeron, J. \& Salpeter, E. 1970;Ap. Letters;7;115

\rf Binney, J. 1977;Ap. J.;215;483

\rf Black, J. 1981;MNRAS;197;553

\rf Blanchard, A., Valls-Gabaud, D. \&
Mamon, G. A. 1992;Astr. Ap.;264;365

\rf Bond, J. R. \& Efstathiou, G. 1984;Ap. J. (Letters);285;L45
  
\rf Bond, J. R. \& Szalay, A. S. 1983;Ap. J.;274;443 

\rf Brainerd, T. G. \& Villumsen, J. V. 1992;Ap. J.;394;409 

\rf Carlberg, R. G. \& Couchman, H. M. P. 1989;Ap. J.;340;47

\rn
Cen, R., Gnedin, N. Y., Koffmann, L. A. \& Ostriker, J. P.
1992, Preprint.
  
\rf Cheney, J. E. \& Rowan-Robinson, M. 1981;MNRAS;195;831

\rf Couchman, H. M. P. \& Rees, M. J. 1986;MNRAS;221;53

\rf Davis, M., Summers, F. J. \& Schlegel, D. 1992;Nature;359;393
  
\rn
Devlin, M. {\etal} 1992, in Proc. NAS Colloquium on Physical
Cosmology, Physics Reports (in press).
 
\rf Efstathiou, G., Bond, J. R. \& White, S. D. M. 1992;MNRAS;258;1P

\rf Efstathiou, G. \& Rees, M. J. 1988;MNRAS;230;5P
 
\rf Efstathiou, G., Frenk, C. S., White, S. D. M.,
Davis, M. 1988;MNRAS;235;715

\rf Feynman, R. P. 1939;Phys. Rev.;56;340

\rn
Gelb, J. M. \& Bertschinger, E. 1992, preprint.

\rn
Gorski, K., Juszkiewicz, R. \& Stompor, R. 1993, preprint.

\rf Gott, J. R. \& Rees, M. J. 1975;Astr. \& Ap.;45;365 
 
\rn
Gundersen, J. O. {\etal} 1993, preprint.

\rn
Jubas, J. \& Dodelson, S. 1993, preprint.

\rf Kashlinsky 1992;Ap.J.;399;L1.

\rn
Klypin, A., Holtzman, J., Primack, J. \& Reg\"os, E. 1993, preprint.

\rf Kompan\'eets, A. 1957;Soviet Phys. -- JETP;4;730

\rf Meinhold, P. R. \& Lubin, P. M. 1991;Ap. J.;370;L11

\rn
Mather {\etal} 1993, preprint.

\rn
Meinhold, P. R. {\etal} 1993, Ap. J. Lett., submitted.

\rf Miralda-Escud\'e, J. \& Ostriker, J. P. 1990;Ap. J.;350;1

\rf O'Brien P.T., Wilson, R \& Gondhalekar, P. M 1988;MNRAS;233;801

\rn Osterbrock, D. E. 1974, {\it Astrophysics of Gaseous Nebulae}
(Freeman, San Francisco).
 
\rf Press, W. H., \& Schechter, P. 1974;Ap. J.;187;425
 
\rf Rees, M. J., \& Ostriker, J. P. 1977;MNRAS;179;541

\rf Rees, M. J.1986;MNRAS;218;25P

\rf Seaton, M. 1959;MNRAS;119;84

\rf Shaefer, R. K. \& Shafi, Q. 1992; Nature;359;199

\rf Shafi, Q. \& Stecker, F. W. 1984;Phys. Rev. Lett.;53;1292
 
\rn
Shuster, J. {\etal} 1993, private communication.
 
\rf Silk, J. I. 1977;Ap. J.;211;638
 
\rf Smoot, G. F. {\etal} 1992;Ap. J. (Letters);396;L1-L18

\rn Spitzer, L. 1968, {\it Diffuse Matter in Space} (Wiley, New York).

\rf Stebbins, A., \& Silk, J. 1986;Ap. J.;300;1

\rn
Subramanyan, R. {\etal} 1993, MNRAS (in press).

\rn
Tegmark, M., Silk, J. \& Evrard, A. 1993, Ap. J. (in press).

\rn W. Vacca 1993, private communication.

\rf Vishniac, E. 1987;Ap. J.;322;597

\rf Vittorio, N. \& Silk, J. 1984;Ap. J. (Letters);285;L39

\rf White, S. D. M., \& Rees, M. J. 1986;MNRAS;183;341
 
\rf Zel'dovich, Y., \& Sunyaev, R. 1969;Ap. Space Sci.;4;301

%%%%%%%%%%%%%%%%%%%%%% FIGURES: %%%%%%%%%%%%%%%%%%%%%%%%%

\clearpage

\clearpage

\clearpage

\clearpage

\clearpage

\clearpage

\clearpage

\clearpage

\clearpage

\clearpage

\end{document}